\documentstyle[12pt,thmsa,sw20lart]{article}

\input tcilatex
\QQQ{Language}{
American English
}

\begin{document}

\topmargin 0pt \oddsidemargin 5mm

\setcounter{page}{1}

\hspace{8cm}Preprint YerPhI-1513(13)-98

\begin{quotation}
\hspace{8cm}{} \vspace{2cm}
\end{quotation}

\begin{center}
{\bf ON THE THEORY OF RELATIVISTIC STRONG PLASMA WAVES}\\\vspace {5mm}

{A.G. Khachatryan}\\\vspace{1cm} {\em Yerevan Physics Institute, Alikhanian
Brothers St. 2, Yerevan 375036, Republic of Armenia}\\E-mail:
khachatr@moon.yerphi.am
\end{center}

\vspace {5mm} \centerline{{\bf{Abstract}}}

The influence of motion of ions and electron temperature on nonlinear
one-dimensional plasma waves with velocity close to the speed of light in
vacuum is investigated. It is shown that although the wavebreaking field
weakly depends on mass of ions, the nonlinear relativistic wavelength
essentially changes. The nonlinearity leads to the increase of the strong
plasma wavelength, while the motion of ions leads to the decrease of the
wavelength. Both hydrodynamic approach and kinetic one, based on
Vlasov-Poisson equations, are used to investigate the relativistic strong
plasma waves in a warm plasma. The existence of relativistic solitons in a
thermal plasma is predicted.

\vspace {5mm}PACS numbers: 52.35.Mw, 52.35.Fp \newpage 

\begin{center}
{\bf I. INTRODUCTION}
\end{center}

Strong plasma waves passing through a plasma with phase velocity slightly
smaller than the velocity of light are the subject of great interest during
last two decades. Such waves can be excited in plasma by relativistic
bunches of charged particles or laser pulses. The excited plasma waves can
be used both to accelerate charged particles and to focus charged bunches
[1]. Plasma-based accelerator concepts are currently under intensive
development (see overview in Ref. [2] and numerous references therein).
Accelerating gradients in the plasma wave can reach the values of tens GeV/m
(notice that in conventional linacs accelerating gradients are in order of
tens MeV/m), that is confirmed in recent experiments [2]. Also the focusing
fields can be much more than that reached in conventional focusing magnetic
systems. The acceleration of charged particles by relativistic strong waves
also is considering as a possible mechanism of ultra-high energy (up to 10$%
^{20}$ eV) cosmic ray generation in astrophysical plasma.

In a cold plasma an amplitude of one-dimensional plasma wave is limited by
the wavebreaking field. In nonrelativistic case, when the wave phase
velocity $v_{ph}$ is much less than velocity of light ($v_{ph}\ll c$), the
wavebreaking amplitude is equal to [3] $E_{*}=m_e\omega _{pe}v_{ph}/|e|$,
where $\omega _{pe}=(4\pi n_0e^2/m_e)^{1/2}$ is the electron plasma
frequency, $n_0$ is the density of electrons in unperturbed plasma, $m_e$
and $e$ are the electron rest mass and their charge; the ions assumed to be
immobile. In relativistic case the wavebreaking field is equal to [4] $%
E_{rel}=[2(\gamma -1)]^{1/2}/\beta $, here $\beta =v_{ph}/c$, $\gamma
=(1-\beta ^2)^{-1/2}$ is the relativistic factor and $E_{rel}$ is normalized
on $E_{*}$. The one-dimensional relativistic strong waves (RSW) can be
excited in a plasma by wide relativistic bunches of charged particles or
intensive laser pulses [2] (when $k_pa\gg 1$, where $k_p=\omega _{pe}/v_{ph}$%
, $a$ is the characteristic transverse sizes of bunches or pulses).

Another important characteristic of nonlinear plasma waves is the dependence
of the wavelength on the wave amplitude. Both in linear case and in
nonlinear nonrelativistic one, the plasma wavelength in cold plasma is $%
\lambda _p=2\pi v_{ph}/\omega _{pe}$. In the relativistic nonlinear regime,
when plasma electrons get a relativistic velocity in the process of
oscillations, nonlinear wavelength increases with the amplitude [5-8]. In
the ultra-relativistic case ($\gamma \gg 1$), for the wave amplitude $%
E_{mp}\cong E_{rel}$, the wavelength approximately is equal to [5] $%
4(2\gamma )^{1/2}\lambda _p$. One can see that for large $\gamma $ the
wavelength is essentially more than usual linear plasma wavelength.

In the previous studies the plasma ions, in the process of oscillations,
including the nonlinear relativistic waves, usually were assumed to be
immobile due to their large mass. In Ref. [9] it is shown that when $\gamma
\ll (M/16m_e)^{1/3}$ (here $M$ is mass of an ion; for example, for hydrogen
plasma, consisting of protons and electrons, this condition gives $\gamma <5$%
), the wavebreaking amplitude approximately is equal to $E_{rel}$ and the
motion of ions can be neglected. However, the dispersion properties of the
relativistic strong plasma waves, which take into consideration the motion
of ions, are not elucidated up to now. This problem has been considered in
Sec. II on the base of cold hydrodynamics equations for the arbitrary $%
\gamma $ and mass of particles forming the plasma. The necessity to take
into account the ion motion conditioned by the following reasons. Firstly,
because the maximum relativistic wavelength and amplitude grow proportional
to $\gamma ^{1/2}$, the plasma ions (even heavy ions) in such strong field
can get velocity which is sufficient to give essential contribution in the
process of charge separation in the wave. On the other hand, in a
semiconductor plasma positively-charged particles (holes) have a mass
similar or less than that of electron. The problem has also astrophysical
aspect. The pole region of the pulsars is considered to be filled with an
electron-positron plasma in which the strong plasma waves can be excited and
high energy charged particles generated [10]. It is obvious that in the
plasma wave passing through an electron-positron plasma, neither electrons
nor positrons may be considered as neutralizing background.

Another important problem is the influence of plasma temperature on RSW.
Finite plasma temperature has decisive significance for description of RSW
near ''wavebreaking''. Actually, according to one-dimensional theory of
relativistic plasma waves in a cold plasma, at the ''breaking'' point
hydrodynamic plasma electron velocity is equal to the phase velocity and the
density of electrons $n_e$ tends to infinity [5-7, 11] (in this case spatial
behavior of the density is similar to the $\delta $-function). On the other
hand, in a thermal plasma (even when the temperature is low), the pressure
tends to infinity when $n_e\rightarrow \infty $. Thus, in this case, the
pressure, as well as the plasma temperature, should be taken into
consideration. In Ref. [12] the finite plasma temperature effect on
nonrelativistic ($v_{ph}\ll c$) wavebreaking field is considered using 1D
waterbag model for the distribution function. In this model it is assumed,
that the electron distribution function during oscillations is constant in a
limited interval of velocities and is equal to zero outside of this
interval. It is shown [12] that maximum amplitude of the plasma waves
decreases with the temperature. In Ref. [13] the 1D relativistic waterbag
model is used to investigate RSW in a warm plasma. Using relativistic
equation of motion with the pressure term for plasma electrons, in Ref. [11]
it is shown that in the case $\gamma \gg m_ev_{ph}^2/3T$ (where $T$ is the
temperature of electrons) the wavebreaking field is proportional to $%
T^{-1/4} $. The authors of Ref. [14] have analyzed the influence of low
temperature ($T\ll m_ec^2$) on excitation of nonlinear wake fields by a
relativistic charged bunches. They considered the equation of motion
obtained using second moments of the distribution function. In the present
paper (Sec. III) the hydrodynamics equations are used to study dispersion
properties of RSW in a warm plasma. The dispersion correlation for weakly
nonlinear case is obtained. In Sec. IV the strong plasma waves are
investigated on the base of relativistic Vlasov kinetic equation and the
Poisson equation.

\begin{center}
{\bf II. ION MOTION EFFECT ON DISPERSION PROPERTIES OF RELATIVISTIC STRONG
PLASMA WAVES}
\end{center}

In this section we consider a cold uniform plasma consisting of
positively-charged particles (for example, protons or positrons) with mass $%
m_{+}$ and electric charge $q_{+}$, and negatively-charged particles
(electrons or negatively-charged ions) with mass $m_{-}$ and the charge $%
q_{-}$. The relativistic equation of motion and the continuity equation for
each plasma component and the Poisson equation for one-dimensional steady
plasma waves are:

\begin{equation}
(\beta -\beta _{\pm })\frac{d(\beta _{\pm }\gamma _{\pm })}{dz}=-\frac{%
q_{\pm }}{|q_{-}|}\beta ^2E,  \tag{1}
\end{equation}

\begin{equation}
\beta \frac{dN_{\pm }}{dz}-\frac{d(N_{\pm }\beta _{\pm })}{dz}=0,  \tag{2}
\end{equation}

\begin{equation}
\frac{dE}{dz}=1-N_{-}+|q_{+}/q_{-}|N_{+},  \tag{3}
\end{equation}
where $z=k_p(Z-v_{ph}t)$, $k_p=\omega _p/v_{ph}$, $\omega _p=(4\pi
n_{0-}q_{-}^2/m_{-})^{1/2}$, $n_{0-}$ is the density of negatively-charged
particles in equilibrium, $\beta _{\pm }=v_{\pm }/c$ are dimensionless
velocities, $\gamma _{\pm }=(1-\beta _{\pm }^2)^{-1/2}$, densities $N_{\pm } 
$ are normalized on the equilibrium values. The electric field strength is
normalized on the nonrelativistic wavebreaking field $m_{-}\omega
_pv_{ph}/|q_{-}|$ and obeys the formula

\begin{equation}
E(z)=-(1/\beta ^2)d\Phi /dz,  \tag{4}
\end{equation}
where $\Phi \equiv \Phi _{-}=1+|q_{-}|\varphi /m_{-}c^2\geq 1/\gamma $, $%
\varphi $ is the electric potential. From expressions (1), (2) and (4) we
have:

\begin{equation}
\beta _{\pm }=[\beta -(\Phi _{\pm }^2-\gamma ^{-2})^{1/2}]/(\beta ^2+\Phi
_{\pm }^2),  \tag{5}
\end{equation}

\begin{equation}
N_{\pm }=\beta \gamma ^2[\Phi _{\pm }^{}/(\Phi _{\pm }^2-\gamma
^{-2})^{1/2}-\beta ].  \tag{6}
\end{equation}
Substituting $N_{\pm }(\Phi _{\pm }^{})$ and expression (4) in the Poisson
equation (3) one obtain the following differential equation of the second
order for $\Phi $: 
\begin{equation}
\frac{d^2\Phi }{dz^2}+\beta ^3\gamma ^2[\frac{\Phi _{+}}{(\Phi _{+}^2-\gamma
^{-2})^{1/2}}-\frac \Phi {(\Phi ^2-\gamma ^{-2})^{1/2}}]=0,  \tag{7}
\end{equation}
Here $\Phi _{+}=1-q_{+}\varphi /m_{+}c^2=1+\mu (1-\Phi )\geq 1/\gamma $ and $%
\mu =|q_{+}/q_{-}|m_{-}/m_{+}$. The electric potential $\varphi $ is assumed
to be equal to zero when plasma density is equal to the equilibrium density.

Equation (7) can be rewritten in the form

\begin{eqnarray}
\frac{d^2\Phi }{dz^2}+\frac{dU}{d\Phi } &=&0,  \tag{8} \\
U &=&\beta ^3\gamma ^2\{[\beta -(\Phi _{+}^2-\gamma ^{-2})^{1/2}]/\mu
+[\beta -(\Phi ^2-\gamma ^{-2})^{1/2}]\}.  \nonumber
\end{eqnarray}
Here, for convenience, $U(\Phi )$ is chosen to be equal to zero in a point $%
\Phi =1$, where it reaches a minimum. When $\mu \rightarrow 0$, Eq. (7)
reduces to the known equation for nonlinear waves in a plasma with immobile
ions [2,7]. Formally, Eq. (8) describes one-dimensional motion of a particle
in a field with potential $U(\Phi )$; the values $\Phi $ and $E$ correspond
to the coordinate and velocity of this fictitious particle respectively.
Function $U$ determines the characteristic of the field moving through the
plasma. In Fig. 1 this function is presented for $\gamma =10$ $(\beta
\approx 0.995)$ and different values of $\mu $. One can see that for the
arbitrary parameters the solutions of Eq. (8) [or Eq. (7)] are the periodic
plasma waves (including the wave with zero amplitude-unperturbed plasma).
Integrating Eq. (8) we have

\begin{equation}
\frac{d\Phi }{dz}=-\beta ^2E=\pm [2(U_{\max }-U)]^{1/2},  \tag{9}
\end{equation}
where $U_{\max }$ is maximum value of $U(\Phi )$ in the process of
oscillations. From (9) it follows that the plasma wave amplitude is equal to 
$E_{mp}=(2U_{\max })^{1/2}/\beta ^2$. Substituting maximum permissible value
of $U(\Phi )$, which reaching at $\Phi =1/\gamma $, in this expression, we
find the wavebreaking field:

\begin{eqnarray}
E_{WB} &=&2^{1/2}\gamma [1+(1-\xi _1^{1/2}\xi _2^{1/2})/\mu ],  \tag{10} \\
\xi _1 &=&1+\mu ,\text{ }\xi _2=1+\mu (\gamma -1)/(\gamma +1).  \nonumber
\end{eqnarray}
In the case $\mu \ll 1$, from (10) follows the expression

\begin{equation}
E_{WB}\approx (1+\mu /8)[2(\gamma -1)]^{1/2}/\beta ,  \tag{11}
\end{equation}
which reduces to the well known relativistic wavebreaking field for plasma
with immobile ions, when $\mu =0$ [4]. Fig. 2 shows the wavebreaking field $%
E_{WB}$ depending on $\mu $ (for example, for electron-positron plasma $\mu
=m_e/m_{pos}=1$, for the hydrogen plasma $\mu =m_e/m_{prot}\approx
5.455\times 10^{-4}$) for different values of $\gamma $. Both in
nonrelativistic case and in relativistic one the wavebreaking field weakly
increases with $\mu $. For example, according to (10), in nonrelativistic
case ($\gamma \approx 1$), $E_{WB}(\mu =1)$ only $2(1-2^{-1/2})^{1/2}\approx
1.08$ times exceeds the wavebreaking amplitude at $\mu =0$. Proceeding from
the shape of the ''potential'' $U(\Phi )$ (see Fig. 1) one can expect that
the plasma wavelength undergoes considerable change with $\mu $. In Fig. 3
the dependence of relativistic plasma wavelength $\Lambda _p$ [note, that
according to the variables accepted in (1)-(3), the linear plasma wave at $%
\mu =0$ corresponds to the value $\Lambda _p=2\pi $] on amplitude presented.
The curve 1 corresponds to the case of immobile ions and coincides with that
previously obtained [7]; in this case $\Lambda _p$ grows with the amplitude
due to relativistic velocities of the oscillating plasma particles (notice,
that in nonlinear nonrelativistic regime the plasma wavelength does not
depend on amplitude). The motion of positive ions for fixed amplitude causes
the decrease of charge separation length and therefore, leads to the
decrease of the wavelength. Fig. 3 clearly shows competition of two
tendencies. For the small $\mu $ the wavelength grows with the amplitude due
to nonlinearity. With the increase of $E_{mp}$, the effect of ion motion
becomes more essential. When $\mu $ is not small, the behavior of the
wavelength caused mainly by ion motion. In electron-positron plasma $(\mu
=1) $, $\Lambda _p$ monotonously decreases with the increase of $E_{mp}$, in
contrast to the case of heavy ions $(\mu \approx 0)$. The results of
simulations presented in Fig. 3 conform with the well known result of linear
theory ($E_{mp}\ll 1$; see, e.g., Ref. [15]): $\Lambda _p=\Lambda
_{p0}/(1+\mu )^{1/2}$, where $\Lambda _{p0}=\Lambda _p(\mu =0)$. Note also
that the results practically did not change for arbitrary $\gamma \gg 1$.
Considerable decrease of the relativistic plasma wavelength with $\mu $ is
demonstrated in Fig. 4.

Previous studies have shown that energy of electrons (or positrons)
accelerated in the field of relativistic nonlinear wave, in a cold plasma
with immobile ions, can reach a value of $4m_ec^2\gamma ^3$ [7,16]. In the
general case the relativistic factor of a resonant electron passing from a
point with the dimensionless potential $\Phi _1$ to a point with $\Phi _2$
is equal to [7]

\begin{equation}
\gamma _{acc}\approx \gamma _{acc}(0)+2\gamma ^2(\Phi _2-\Phi _1),  \tag{12}
\end{equation}
where $\gamma _{acc}(0)\approx \gamma $ is value of the relativistic factor
at the initial point. When $\mu =0$, $\Phi _{\min }=1/\gamma \leq \Phi \leq
\Phi _{\max }\approx 2\gamma $ [7]. Substituting this maximum and minimum
values in (12), for maximum energy of accelerated electrons one obtains $%
(\gamma _{acc})_{\max }\approx 4\gamma ^3$. With $\mu $ growth, the maximum
energy decreases due to the decrease of $\Phi _{\max }$ (see Fig. 1). For
the case of electron-positron plasma function $U(\Phi )$ is symmetric with
reference to axis $\Phi =1$ and, as it is easy to see, in this case $\Phi
_{\min }=1/\gamma $, $\Phi _{\max }=2-1/\gamma $. Then, the maximum energy
of accelerated particles is $(\gamma _{acc})_{\max }\approx 4\gamma ^2$,
that is in $\gamma $ times less than that in the case $\mu =0$.

If a bunch of charged particles with density $n_b(z)$ and electric charge $%
q_b$ passes through a plasma, adding to the left side of equation (7) the
value $\alpha (z)=\beta ^2(q_b/|q_{-}|)n_b(z)/n_{0-}$, we obtain the
equation that describes the excitation of steady plasma wake fields by the
bunch. In this case the phase velocity is equal to velocity of the bunch. In
this section the properties of RSW have been investigated by simulation of
wake wave generation by charged bunches.

Above we have considered the case $\mu \leq 1$. However, one can see that
the obtained results are valid also for $\mu >1$, if $E$ we replace by $-E$
[this new $E$ is normalized to $m_{+}\omega _pv_{ph}/q_{+}$, $\omega
_p=(4\pi n_{0+}q_{+}^2/m_{+})^{1/2}$], $\mu $ replace by $1/\mu $, replace
the subscript $"+"$ by $"-"$ and vice versa.

\begin{center}
{\bf III. THE INFLUENCE OF ELECTRON TEMPERATURE ON RELATIVISTIC NONLINEAR
PLASMA WAVES: HYDRODYNAMIC APPROACH}
\end{center}

Here we continue to consider the dispersion properties of RSW in the frame
of hydrodynamic approach, investigate relativistic nonlinear waves in a warm
plasma. Adding the relativistic pressure term $-(\gamma _e^2/N_e)(1-\beta
\beta _e)dP/dz$ [11,14] with $P=\tau (N_e/\gamma _e)^3$ [11] (which is
relativistic generalization of usual equation of state for one-dimensional
adiabatic compression) to the equation of motion of plasma electrons one can
obtain the equations:

\begin{eqnarray}
(\beta -\beta _e)\frac{d(\beta _e\gamma _e)}{dz} &=&\beta ^2E+3\beta ^2\tau 
\frac{\gamma _e^{}(1-\beta \beta _e)^2}{(\beta -\beta _e)^3}\frac{d\beta _e}{%
dz},  \tag{13} \\
\frac{dE}{dz} &=&-\frac 1{\beta ^2}\frac{d^2\Phi }{dz^2}=1-N_e.  \nonumber
\end{eqnarray}
In (13) $\beta _e=v_e/c$ and $\tau =T/m_ec^2$ are dimensionless velocity and
temperature of plasma electrons, $\gamma _e=(1-\beta _e^2)^{-1/2}$, $\Phi
=1+|e|\varphi /m_ec^2$. The plasma ions are assumed to be immobile due to
their large mass. The density of electrons $N_e=n_e/n_0$ normalized to the
unperturbed value $n_0$, as usually, is obtained from the continuity
equation:

\begin{equation}
N_e=\beta /(\beta -\beta _e).  \tag{14}
\end{equation}
When $\tau \rightarrow 0$ equations (13) and (14) describe RSW in a cold
plasma [5,6]. Note, that the value $\tau =1$ corresponds to the temperature
of about 6$\times $10$^9$ K. For laboratory plasmas the temperature changes
in the bounds $\tau \sim 10^{-6}\div 10^{-2}$; for star plasmas $\tau \sim
10^{-5}-1$.

The dispersion correlation can be obtained analytically for weakly nonlinear
wave, when $u=\beta _e(z)/\beta \ll 1$. In this case from equations (13) and
(14) we have:

\begin{equation}
(a_0-a_1u+a_2u^2)\frac{d^2u}{dz^2}-(a_1-2a_2u)\left( \frac{du}{dz}\right)
^2+u+u^2+u^3=0,  \tag{15}
\end{equation}

\[
a_0=1-3\tau /\beta ^2,a_1=1+3\tau (3-2\beta ^2)/\beta ^2,a_2=3\beta
^2/2-3\tau (6-11\beta ^2/2+\beta ^4)/\beta ^2. 
\]
Looking for solution of Eq. (15) as (see, e.g., Ref. [17])

\begin{eqnarray*}
u &=&\varepsilon u_1(\Psi )+\varepsilon ^2u_2(\Psi )+\varepsilon ^3u_3(\Psi
)+..., \\
d\Psi /dz &=&\lambda _p/\Lambda _p=\kappa _0+\varepsilon \kappa
_1+\varepsilon ^2\kappa _2+...,
\end{eqnarray*}
we obtain

\begin{eqnarray}
\lambda _p/\Lambda _p &=&a_0^{-1/2}(1+b\beta _m^2),  \tag{16} \\
b &=&-3/16+3\tau a_0^{-1}(10-9\beta ^2+\beta ^4)/8\beta ^4+3\tau
^2a_0^{-2}(2-\beta ^2)^2/\beta ^6,  \nonumber
\end{eqnarray}
where $\varepsilon =\beta _m/\beta \ll 1$ is the small parameter, $\beta
_m=(\beta _e)_{\max }$ and $\Lambda _p$ is the wavelength. In the linear
case ($\beta _m^2\rightarrow 0$) from (16) it follows that the wavelength
decreases with the temperature according to well known Bohm-Gross dispersion
correlation: $\Lambda _p=\lambda _p(1-3\tau /\beta ^2)^{1/2}$. On the other
hand, in a cold plasma ($\tau =0$) the wavelength increases due to
nonlinearity (see, e.g., Ref. [15]): $\Lambda _p\approx \lambda _p(1+3\beta
_m^2/16)$. Fig. 5 shows the dependence of the wavelength on wave amplitude
in thermal plasma obtained by simulation of Eqs. (13) and (14). In the case
of low temperature this dependence almost coincides with that in a cold
plasma (compare curves 1 in Figs. 5 and 3).

In the case $\beta \rightarrow 1$ equations (13) and (14) can be easily
integrated (see also Ref. [11]):

\begin{equation}
\Phi -\left( \frac{1-\beta _e}{1+\beta _e}\right) ^{1/2}-3\tau \left[ \left( 
\frac{1+\beta _e}{1-\beta _e}\right) ^{1/2}-1\right] =0.  \tag{17}
\end{equation}
One can see that the thermal term can not be neglected near the
''wavebreaking'' ($\beta _e\rightarrow 1$) even for low temperature. In the
latter case the wavebreaking field is proportional to $\tau ^{-1/4}$ [11].

In the frame of hydrodynamic theory the velocity of plasma electrons can not
exceed the wave phase velocity. Actually, if $\beta _e>\beta $, then,
according to expression (14), the density of electrons becomes negative,
that has no physical sense. In reality, when $\beta _e\approx \beta $ in a
warm plasma (even when the temperature is low), considerable part of
electrons gets velocities more than the phase velocity due to their thermal
energy distribution. When the temperature is not low, the energy
distribution effect on plasma waves is essential in all cases. Therefore,
the strong waves near the ''wavebreaking'' and when $\tau $ is not low, can
be described correctly in the frame of kinetic approach.

\begin{center}
{\bf IV. KINETIC THEORY OF THE RELATIVISTIC STRONG PLASMA WAVES}
\end{center}

As in the previous section, here we assume plasma ions to be immobile. The
kinetic approach, for one-dimensional steady fields passing through a warm
plasma, gives the following system, obtained from relativistic Vlasov
equation and Maxwell equations (see, e.g., Ref. [18]):

\begin{equation}
\left[ \beta -\frac p{(1+p^2)^{1/2}}\right] \frac{\partial f}{\partial z}-%
\frac{\partial \Phi }{\partial z}\frac{\partial f}{\partial p}=0,  \tag{18}
\end{equation}

\begin{equation}
\frac{d^2\Phi }{dz^2}+\beta ^2(1-N_e)=0,  \tag{19}
\end{equation}

\begin{equation}
N_e=\int_{-\infty }^{+\infty }f(p,z)dp,  \tag{20}
\end{equation}
where $p=p_z$ is the plasma electron momentum, normalized to $m_ec$, $f(p,z)$
is the distribution function. In an unperturbed plasma the distribution
function is equal to 1D relativistic Maxwell distribution [19]

\begin{eqnarray}
f_0 &=&A[1+(1+p^2)^{1/2}/\tau ]\exp [-(1+p^2)^{1/2}/\tau ],  \tag{21} \\
A &=&\tau /2K_2(1/\tau ),  \nonumber \\
\langle p^2\rangle _0 &=&\tau [K_1(1/\tau )/K_2(1/\tau )+4\tau ],  \nonumber
\end{eqnarray}

\[
\langle (1+p^2)^{1/2}\rangle _0=2\tau +(1-\tau ^2)K_1(1/\tau )/K_2(1/\tau ), 
\]
where $K_n(x)$ is the modified Bessel function of the $n$-th order. In (21)
we also have written out average squared pulse and total energy for the
one-dimensional equilibrium distribution $f_0$, that may be interesting for
future investigations.

Solving Eq. (18) by the method of characteristics (see, e.g., Ref. [20]) and
requiring that the function $f$ reduces to the equilibrium distribution (21)
at $\Phi =1$, one obtains the following general solution:

\begin{eqnarray}
f &=&A(1+S/\tau )\exp (-S/\tau ),  \tag{22} \\
S &=&(1+g^2)^{1/2},\text{ }g=-\gamma ^2[\beta r\pm (r^2-\gamma ^{-2})^{1/2}],
\nonumber \\
r &=&\beta p-(1+p^2)^{1/2}+\Phi -1,  \nonumber
\end{eqnarray}
In the expression for $g$, the plus sign corresponds to the case $p\leq
\beta \gamma $ and the minus sign to $p>\beta \gamma $; in equilibrium ($%
\Phi =1$) we have $g=p$. Substituting the expressions (20) and (22) in Eq.
(19), one obtains equation for $\Phi $. The plasma electron density $%
N_e(\Phi )$ obtained numerically from (20) and (22) is presented in Fig. 6.
In the case of low temperature, the integral in (20) can be calculated by
the Laplace asymptotic method [19]. The value of $N_e$ is at maximum when $%
\Phi \approx 1/\gamma $ and is equal to

\begin{equation}
N_{\max }\approx [\Gamma (1/4)/4]\gamma (\beta \gamma /\pi )^{1/2}(2/\tau
)^{1/4}\approx 0.6\gamma (\beta \gamma )^{1/2}\tau ^{-1/4}.  \tag{23}
\end{equation}
According to expression (23), in a cold plasma $N_{\max }\rightarrow \infty $%
, that conforms with the previous investigations [5,6,11]. When $\tau \ll 1$
and $\Phi >1/\gamma $, the dependence $N_e(\Phi )$ approximately is
described by expression (6).

Simulations of the problem show that the plasma wavelength increases with
the wave amplitude and tends to infinity for a solitary wave (soliton). Fig.
7 shows dependence of the maximum value of the amplitude (which corresponds
to the maximum of electric field strength in the soliton $E_s$) on plasma
temperature for different $\gamma $. One can see that $E_s$ is almost
constant and equal to the relativistic wavebreaking field $E_{rel}$ for the
values of $\tau $ up to $0.05-0.1$ and then decreases rapidly.

As it was mentioned above, the relativistic plasma waves can be excited by
charged bunches or laser pulses. In order to describe the excitation of the
wake field by a charged bunch, it is necessary to add in the left side of
Eq. (19) the value $\beta ^2\alpha (z)$ [the definition of $\alpha (z)$ see
in Sec. II]. The nonlinear periodical wave and the solitary wave excited by
uniform electron bunch are plotted in Fig. 8. The plasma electron density
behind the soliton tends to its equilibrium value ($N_e\rightarrow 1$) and
the electric field strength tends to zero. However, in this case the average
plasma electron momentum $\langle p\rangle =\int_{-\infty }^{+\infty
}pf(p,z)dp/\int_{-\infty }^{+\infty }f(p,z)dp$ tends to a non-zero constant
value. This does not seem strange because the solitary wave can be
considered as a wave with infinite wavelength. Hence, bulk motion of plasma
electrons behind the solitary wave takes place, while the plasma remains
neutral. Equations (18)-(21) have also non-periodical solutions. However,
such solutions have no physical sense [19] and should be considered in the
frame of non-stationary kinetic theory. Thus, in a thermal plasma with
immobile ions two kinds of steady waves can exist: periodical waves and
solitons.

\begin{center}
{\bf V. CONCLUSIONS}
\end{center}

The results presented in this paper supplement the theory of nonlinear
relativistic plasma waves, taking into consideration motion of ions and
finite plasma temperature. It is shown that the nonlinearity leads to the
increase of relativistic wavelength, while ion motion leads to the
wavelength decrease. For example, in electron-positron plasma the wavelength
monotonously decreases as the amplitude increases. The relativistic
wavebreaking field weakly depends on the ion mass.

Contrary to the case of cold plasma, in a warm plasma the relativistic
solitary waves (solitons) can exist. The plasma wavelength grows
monotonously with the amplitude in the warm plasma due to nonlinearity. It
has been found that maximum electron density in the plasma wave decreases
with the temperature as $T^{-1/4}$ and tends to infinity in a cold plasma,
that was shown by previous investigations.

\begin{center}
{\bf ACKNOWLEDGMENTS}
\end{center}

The author would like to thank S. S. Elbakian and E. V. Sekhpossian for
helpful discussions and H. Nersisyan for the help in preparing the
manuscript for publication.

This work was supported by the International Science and Technology Center,
Project A-013.

\begin{center}
{\bf REFERENCES}
\end{center}

[1] P. Chen, Part. Accel. {\bf 20}, 171 (1987).

[2] E. Esarey, P. Sprangle, J. Krall, and A. Ting, IEEE Trans. Plasma Sci. 
{\bf 24}, 252 (1996).

[3] J. M. Dawson, Phys. Rev. {\bf 133}, 383 (1959).

[4] A. I. Akhiezer and R. V. Polovin, Sov. Phys. JETP {\bf 3}, 696 (1956).

[5] A. Ts. Amatuni, E. V. Sekhpossian, and S. S. Elbakian, Fiz. Plazmy {\bf %
12}, 1145 (1986).

[6] J. B. Rosenzweig, Phys. Rev. Lett. {\bf 58}, 555 (1987).

[7] A. G. Khachatryan, Phys. Plasmas {\bf 4}, 4136 (1997).

[8] K. Nakajima, Phys. Rev. A {\bf 45}, 1149 (1992).

[9] A. Ts. Amatuni, E. V. Sekhpossian, and S. S. Elbakian, Preprint
YerPhI-1176(53)-89, Yerevan, 1989.

[10] E. Asseo, X. Llobet, and G. Schmidt, Phys. Rev. A {\bf 12}, 1293 (1980).

[11] J. B. Rosenzweig, Phys. Rev. A {\bf 38}, 3634 (1988).

[12] T. P. Coffey, Phys. Fluids {\bf 14}, 1402 (1971).

[13] T. Katsouleas and W. B. Mori, Phys. Rev. Lett. {\bf 61}, 90, (1988).

[14] A. Ts. Amatuni, E. V. Sekhpossian, and S. S. Elbakian, Part. Accel. 
{\bf 36}, 241 (1992).

[15] P. K. Shukla, N. N. Rao, M. Y. Yu, and N. L. Tsintsadze, Physics
Reports {\bf 138}, N 1\&2 (1986).

[16] E. Esarey and M. Pilloff, Phys. Plasmas {\bf 2}, 1432 (1995).

[17] G. B. Witham, {\it Linear and Nonlinear Waves} (John Wiley \& Sons, New
York, 1974), chap. 13.

[18] S. R. de Groot, W. A. van Leeuven, and Ch. G. van Weert, {\it %
Relativistic Kinetic Theory} (North-Holland, Amsterdam, 1980).

[19] A. G. Khachatryan, Phys. Plasmas {\bf 5}, 112 (1998).

[20] G. A. Korn and T. M. Korn, {\it Mathematical Handbook} (McGraw-Hill,
New York, 1968), chap. 10.

\end{document}